\documentclass[a4paper,11pt]{article}
\pdfoutput=1
\usepackage{jheppub}
\usepackage[T1]{fontenc}
\usepackage{float}
\usepackage{marginnote}
\usepackage{comment}

\def\Xint#1{\mathchoice
   {\XXint\displaystyle\textstyle{#1}}%
   {\XXint\textstyle\scriptstyle{#1}}%
   {\XXint\scriptstyle\scriptscriptstyle{#1}}%
   {\XXint\scriptscriptstyle\scriptscriptstyle{#1}}%
   \!\int}
\def\XXint#1#2#3{{\setbox0=\hbox{$#1{#2#3}{\int}$}
     \vcenter{\hbox{$#2#3$}}\kern-.5\wd0}}

\def\dashint{\Xint-}

\def\ep{\varepsilon}
\def\half{\frac{1}{2}}

\def\K{\mathcal{K}}
\def\F{\mathcal{F}}

\def\lr#1{\left(#1\right)}
\def\slr#1{\left[#1\right]}

\def\trl#1{\textrm{Tr}\lr{#1}}

\def\avg#1{\left\langle #1\right\rangle}

\def\CPn{\mathbb C P^n}

\def\RS2{\mathbb R\times S^2_F}

\def \be  {\begin{equation}}
\def \ee  {\end{equation}}
\def \bex  {\begin{equation*}}
\def \eex  {\end{equation*}}
\def \bea {\begin{eqnarray}}
\def \eea {\end{eqnarray}}
\def \bal {\begin{align}}
\def \eal {\end{align}}

\def\no{\nonumber\\}

\def \PRD {{Phys. Rev. D\ }}
\def \JHEP {{JHEP\ }}

\allowdisplaybreaks
\title{\boldmath Second moment fuzzy-field-theory-like matrix models}

\author{M\'{a}ria \v{S}ubjakov\'{a}}
\author{Juraj Tekel}
\affiliation{Department of Theoretical Physics\\ Faculty of Mathematics, Physics and Informatics, Comenius University\\ Mlynsk\'a Dolina, Bratislava, 842 48, Slovakia
}
\emailAdd{maria.subjakova@fmph.uniba.sk}
\emailAdd{juraj.tekel@fmph.uniba.sk}

\abstract{
We solve a multitrace matrix model approximating the real quartic scalar field theory on the fuzzy sphere and obtain its phase diagram. We generalize this method to models with modified kinetic terms and demonstrate its use by investigating models related to the removal of the UV/IR mixing. We show that for the fuzzy sphere a modification of the kinetic part of the action by higher derivative term can change the phase diagram of the theory such that the triple point moves further from the origin.
}

\begin{document} 
\maketitle
\flushbottom

\section{Introduction}

Spaces with noncommuting coordinates have been a part of theoretical physics for quite some time. As a fundamental concept \cite{snyder,sF22}, as an effective description of different phenomena \cite{qhe2,witten}, or as various solutions and backgrounds in matrix model descriptions of string theory \cite{br1,br2}.

Fuzzy spaces are finite mode approximations to compact manifolds \cite{steinacker_review}. The space is divided into a finite number of cells, not unlike the phase space of quantum mechanics. As such, the field theories on fuzzy spaces have finite number of degrees of freedom and are essentially matrix models. These properties make the fuzzy spaces a very important setting to test the consequences of quantum structure of spacetimes, which is expected to be present in the quantum theory of gravity \cite{doplicher}.

Any matrix model begs to be put on a computer. Matrix models describing the fuzzy field theories have been investigated in numerous Monte Carlo studies: for the fuzzy sphere \cite{num09,num14,OConSamo}, for the fuzzy disc \cite{num_disc}, for the fuzzy sphere with a commutative time, i.e. the three dimensional space ${\mathbb R \times S_F^2}$ \cite{num_RSF2}, for the fuzzy torus \cite{num14panero2} and for field on the noncommutative plane coupled to a curvature term \cite{srbski}.\footnote{See \cite{panero15} for a review of numerical investigations into noncommutative field theories.} All these works point to existence of a noncommutative phase of the theory, which breaks the translation invariance of the underlying space. Here the field is organized into striped regions of oscillations around different minima of the potential and the existence of this phase has been established computationally even earlier \cite{NCphase1}. It exists together with the two standard field theory phases, the symmetric disorder phase and the uniform order phase breaking the $\phi\to-\phi$ symmetry of the action. The phase transition lines between these phases meet at a triple point. More recently, properties of the correlation functions \cite{correlationFunctions1,correlationFunctions2} and the entanglement entropy \cite{entanglement1,entanglement2,entanglement3} on the fuzzy sphere have been investigated numerically.

The analytical treatment of the corresponding matrix models is complicated. The problem is, as we will shortly see, that the part of the probability distribution which comes from the kinetic term of the field theory breaks the unitary symmetry. Thus the standard procedure of diagonalization is no longer straightforward. There have been several different approaches to circumvent this issue.

One is to consider the kinetic term as a perturbation \cite{OConSaman}. After the expansion of the probability distribution in powers of the kinetic term, it is possible to perform the unitary integral and after re-exponentiation, one is left with a complicated multi-trace model for the eigenvalues. This was first done for the fuzzy sphere \cite{OConSaman,samann,samann2} and similar analysis has been performed also the three dimensional ${\mathbb R \times S_F^2}$ \cite{samanRfuzzyS} and for the fuzzy disc \cite{samanfuzzydisc}. The model is well behaved for large values of the parameters, but fails close to the origin of the parameter space, where the triple point is located.

The second approach \cite{poly13} is based on the fact that for the free field theory (i.e. without interaction) the matrix model can be solved exactly \cite{steinacker05,PNT12}. One can then reorganize the multitrace action terms into a group that vanishes for the solution of the free model and terms depending only on the symmetrized second moment. The second contribution can be computed and the first group can be dropped as an approximation.

In this work, we continue the line of research for the fuzzy sphere started in \cite{jt18}, where the equations describing the second moment matrix model have been solved numerically, the phase diagram with all the three phases has been obtained and the triple point has been identified. The location agreed qualitatively with the previous Monte Carlo simulations \cite{OConSamo}.

Here, we solve the equations analytically. Equations themselves are transcendental, but can be solved order by order in perturbative expansions. We then complete the solution using the method of Pade approximants and reconstruct the results of our previous work, however with a firm analytical hold and control. We then generalize the approach to different forms of the kinetic term and work out results for modification of the kinetic term on the fuzzy sphere.

After giving some very basic preliminary information in the Section \ref{sec2} we proceed to the solution of the second moment matrix model for the fuzzy sphere in the Section \ref{sec3}. We obtain all three phase transition lines of the model and then analyze the location of the triple point in the Section \ref{sec3.4}. In the first part of the Section \ref{sec4}, we outline a solution to a model given by a general effective kinetic term action. In the rest of the section, we apply this method to two models motivated by the fuzzy field theory with no UV/IR mixing.

\section{Preliminaries}\label{sec2}
In this section we provide some essential preliminaries for what follows in further sections. To keep this report as short as possible, we concentrate on the notions and expression we will directly use. A more thorough review of the topics of matrix models and fuzzy field theory can be found in \cite{bal,lecturesydri,matrixmodels,jt15b,msjt20,new20}.

\subsection{Matrix models of fuzzy field theories}

The real scalar field on a fuzzy space \cite{bal,lecturesydri} is given by a ${N\times N}$ hermitian matrix $M$, the action
\be\label{2fuzzyact}
S(M)=\trl{\half M \K M+\half r M^2+ V(M)}\ ,
\ee
and the expectation values are given by the functional correlation functions
\be\label{sec2_matrix}
\avg{\mathcal{O}(M)}=\frac{1}{Z}\int dM\,e^{-N^2 S(M)}\,\mathcal{O}(M)\ \textrm{, with}\ Z=\int dM\,e^{-N^2 S(M)}\ .
\ee
The matrix $M$ and parameters of the theory ${r,g}$ are assumed to be rescaled in such a way that any volume factors are absorbed and the $N^2$ scaling in the probability distribution is reproduced.\footnote{One technically rescales the eigenvalues after the matrix is diagonalized.}

As we can see (\ref{sec2_matrix}) is a particular case of a random matrix model \cite{steinacker05,PNT12}. Compared to standard matrix models the model above is complicated by the presence of the kinetic term. At the moment it is not possible to treat this term completely and we will use an approximation first used in \cite{poly13}.\footnote{See \cite{gw19} for the most recent developments.} The kinetic term is captured by an effective action $F$ and a remainder term
\be\label{4model}
S(M)=\half F(c_2-c_1^2)+\mathcal R + \half r c_2+g c_4\ ,\ c_n=\frac{1}{N}\trl{M^n}\ ,
\ee
where from now on we consider the quartic potential. The remainder term $\mathcal R$ vanishes when evaluated for the solution of the free model and we will drop this contribution as an approximation to (\ref{2fuzzyact}). This means that our results will be the more accurate, the closer the configuration of eigenvalues of $M$ is to the $g=0$ case, i.e. the semicircle. For configurations far from the semicircle the approximation provides us with a solvable model but might, and as we will see sometimes will, lead to inaccurate results. 

The model (\ref{4model}) is a multitrace matrix model and models of this form will be referred to as second moment fuzzy-field-theory-like matrix models.\footnote{Recall that in the large $N$ limit such a model is equivalent to a single trace model with the effective parameter ${r_{eff}=r+F'(c_2-c_1^2)}$, eventually determined by the self-consistency conditions on the moments of the distribution \cite{jt15b}.}

\subsection{Solutions of the second moment multitrace models}

The action (\ref{4model}) is invariant under conjugation ${M\to UMU^\dagger}$, the matrix $M$ can be diagonalized and the situation turned into an eigenvalue problem. We are going to be interested in the large $N$ solution and in this case only the saddle point configurations of the eigenvalues contribute to the integral (\ref{sec2_matrix}). In this limit the eigenvalues will form a continuous distribution ${\rho(x)}$ supported on one or more intervals. The saddle point equation for this model has three different types of solutions \cite{japonec} determined by nature of the support of the eigenvalue distribution, see \cite{jt18,msjt20} for details.

If more than one solution for the eigenvalue distribution exists, the preferred solution will be the one with the lower free energy\footnote{The last term in this expression comes from the Vandermonde determinant, which is a jacobian of the transition to the eigenvalue description. The integral $\dashint $ is a principal value integral.}
\be
\F=\half F(c_2-c_1^2)+\half r c_2+g c_4-\dashint dx dy\, \rho(x)\rho(y)\log|x-y|\ ,\ c_n=\int dx\,\rho(x) x^n\ .
\ee

\emph{The symmetric one-cut solution}\\
In this case, the eigenvalue distribution is supported on one symmetric interval ${(-\sqrt \delta,\sqrt \delta)}$, which is determined by the following condition
\begin{align}
0\,=\,&\frac{4-3 \delta^2 g}{\delta}-r - F'\lr{\frac{4\delta + \delta^3 g}{16}}\ .\label{4_cond1}
\end{align}
The free energy is then given by
\begin{align}
\F_{S1C}\,=\,&\frac{1}{4}+\frac{9}{128}\delta^4 g^2+\frac{1}{8}r \delta+\frac{1}{32} \delta^3 g r+\frac{1}{2} F\lr{\frac{4\delta+\delta^3 g}{16}}-\frac{1}{2} \log\left(\frac{\delta}{4}\right)\ .\label{4_freeS1C}
\end{align}
Since the eigenvalues of the matrix represent the values of the field, this solution corresponds to the disorder phase of the field theory.

\emph{The symmetric two-cut solution}\\
In this regime, the eigenvalue density is supported intervals $(-\sqrt{D+\delta},-\sqrt{D-\delta})$ and ${(\sqrt{D-\delta},\sqrt{D+\delta})}$. The endpoints of the two intervals are determined by the conditions
\begin{align}
0\,=\,&4D g+r+F'\lr{D}\ ,\ \delta^2=\frac{1}{g}\ .\label{4_cond2}
\end{align}
The free energy is given by
\begin{align}
\F_{2C}\,=\,&\frac{3}{8}+D^2 g+\frac{D r}{2}+\frac{1}{2}F\lr{D}+\frac{1}{4} \log\lr{4 g}\label{4_freeS2C}\ .
\end{align}
This solution corresponds to the striped phase of the fuzzy field theory.

\emph{The asymmetric one-cut solution}\\
The model has also an asymmetric solution, supported on an interval ${(D-\sqrt \delta,D+\sqrt \delta)}$. $\delta$ is determined by
\be
4\frac{4+15\delta^2 g + 2 r \delta}{\delta(4+9\delta^2 g)}-F'\lr{\frac{\delta \lr{64  + 160 \delta^2 g+144 \delta^4 g^2+81 \delta^6 g^3+36 \delta^3 g r + 27 \delta^5 g^2 r}}{64(4+9\delta^2 g)}}=0\ .\label{4_cond3}
\ee
$D$ is then given by
\be
D^2=\frac{-r(4 +3 \delta^2 g )-12 \delta g-9 \delta^3 g^2}{4 g \left(4+9 \delta^2 g\right)}\ .
\ee
Finally, the formula for the free energy is a complicated and uneventful
\begin{align}
\F_{AS1C}\,=\,&-\frac{1}{128 g \left(4+9 \delta^2 g\right)^2}\Big[
6075 \delta^8 g^5
+3240 \delta^6 g^4 (4+r \delta)
+144 \delta^4 g^3 \left(29+40 r \delta+3 \delta^2 r^2\right)\no
+8 \delta^2 &g^2 \left(-144+352 r \delta+117 \delta^2 r^2\right)+64 g \left(-8+8 r \delta+9 \delta^2 r^2\right)
+128 r^2\Big]+\no
+\half F&\lr{\frac{\delta \lr{64  + 160 \delta^2 g+144 \delta^4 g^2+81 \delta^6 g^3+36 \delta^3 g r + 27 \delta^5 g^2 r}}{64(4+9\delta^2 g)}}-\half \log\lr{\frac{\delta}{4}}\ .\label{4_free1AC}
\end{align}
This solution corresponds to the uniform order phase of the field theory.

\subsection{The fuzzy sphere}
In the case of the fuzzy sphere \cite{madore} the kinetic term is given by
\be
\K M=\frac{1}{R^2}\sum_{i=1}^3[L_i,[L_i,M]]=C_2\,M\ ,
\ee
where $L_i$'s are the generators of $SU(2)$ in the $N$ dimensional representation. We denote the double commutator $C_2$ since it is the quadratic Casimir operator and we will set the radius $R$ of the sphere to $1$ without loss of generality. Following \cite{poly13}, with more details in the Section \ref{sec4.1}, the effective kinetic terms action is
\begin{align}
F(t)=&\log\lr{\frac{t}{1-e^{-t}}}\ , \label{5sphereF}\\
	\stackrel{t\to 0}{=}&\ \frac{1}{2}t-\frac{1}{24}t^2+\frac{1}{2880}t^4+\ldots\label{5sphereF0}\ ,\\
	\stackrel{t\to \infty}{=}&\ \log t+e^{-t}+\frac{1}{2}e^{-2t}+\frac{1}{3}e^{-3t}+\ldots\ .\label{5sphereFinfty}
\end{align}
where we have given also the large and the small $t$ expansions, which are going to be relevant later. These equations have been analyzed numerically \cite{jt18} and the triple point of the model has been located at
\be\label{sec5_triplepoint}
\lr{g_c,r_c}\approx \lr{0.0049,-0.35}\ .
\ee
Here, we will solve the model analytically.

\section{Fuzzy sphere model}\label{sec3}

The form of the effective action for the sphere (\ref{5sphereF}) means that the equations (\ref{4_cond1},\ref{4_cond2},\ref{4_cond3}) become transcendent and there is no chance to solve the equations completely.

A different approach to the numerical solution \cite{jt18} has been employed in an analysis of a similar, yet more simple model in \cite{samanfuzzydisc}. The equations have been solved in the limit of very large parameters $r$ and $g$. This is a very natural thing to try also in our setting. The kinetic term action $F$ acts as an effective coupling and its effect is negligible in the large $r$ limit. We will be able to calculate consecutive corrections to the solution of the matrix model without the kinetic term contribution in powers of $1/r$.

This approach will however not work for the symmetric one-cut solution, since this solution does not exist for $r$ less than $-4\sqrt g$. A different way to see this is to look at the equations for the solutions directly. Taking the negative and large $r$ limit in (\ref{4_cond3}), i.e. small $\delta$ limit, lets us to work with a small parameter expansion of the function ${F(t)}$. The same $r$ limit, i.e. the large $D$ limit in (\ref{4_cond2}) yields a large parameter expansion of ${F(t)}$. However no such limit works nicely for in (\ref{4_cond1}).

We first analyse the symmetric one-cut to two-cut phase transition, which can be calculated exactly even in this setting. We then present the perturbative calculation of the asymmetric one-cut to two-cut transition, which we then treat to a Pade approximation. We perform a similar procedure, however little more technically involved, with the symmetric one-cut to asymmetric one-cut phase transition. This way, we obtain the full phase diagram.

\subsection{Symmetric one-cut to symmetric two-cut transition}\label{sec3.1}

This part can be solved completely \cite{poly13}. Since the phase transition condition can be expressed explicitly in the terms of $\delta$ and $D$ as ${\delta=D=1/\sqrt{g}}$ we get
\be\label{5_line1}
r=-4\sqrt{g}-F'(1/\sqrt{g})=-5\sqrt{g}-\frac{1}{1-e^{1/\sqrt{g}}}\ .
\ee
Note that the second term in (\ref{5_line1}) is exponentially suppressed for small values of $g$. This is important since the triple point is precisely in this region.

\subsection{Asymmetric one-cut to symmetric two-cut transition}\label{sec3.2}

\paragraph{Asymmetric one-cut solution}
In the asymmetric case we expect all the eigenvalues will collapse to the minimum of the potential as ${r\to-\infty}$ for a fixed $g$. Thus we expect ${\delta\to0}$ in this limit and and since the $F'$ term in (\ref{4_cond3}) does not play any role in the leading order, we look for a solution of in the following form
\be\label{expdelta}
\delta=\frac{\delta_1}{r}+\frac{\delta_2}{r^2}+\ldots\ .
\ee
In the whole text ellipses stand for higher order terms in the corresponding expansion, most often negative and large $r$ expansion. (\ref{expdelta}) means we can work with the small $t$ expansion (\ref{5sphereF0}) in (\ref{4_cond3}). Taking the limit of large and negative $r$ and solving the equation order by order yields
\be 
\delta=-\frac{2}{r}-\frac{1}{2 r^2}-\frac{4+720 g}{24 r^3}-\frac{2+864 g}{32 r^4}+\ldots
\ee
with the expansion the free energy (\ref{4_free1AC})
\begin{align}
\F_{AS1C}=&-\frac{r^2}{16 g}+\frac{1}{2} \log\lr{- r}+\frac{3}{4}+\frac{1}{2} \log 2-\frac{1}{8 r}-\frac{1+120 g}{48 r^2}\nonumber\\&-\frac{1+276 g}{192 r^3}-\frac{3+1400 g+54000 g^2}{1920 r^4}+\ldots\ .\label{freeAS1C}
\end{align}
The order of these expressions is determined by the order of the expansion of ${F(t)}$ we use. More terms of the expansion (\ref{5sphereF0}) yield higher order terms in (\ref{freeAS1C}), however going further in the expansion does not change the lower order terms in (\ref{freeAS1C}).

\paragraph{Symmetric two-cut solution}

In the symmetric two-cut case, we expect the two cuts to be localized around the two minima of the potential, i.e. we look for the solution in the form
\be\label{solD}
D=-\frac{r}{4g}+D_0+\frac{D_1}{r}+\frac{D_2}{r^2}+\ldots\ .
\ee
This means we can work with the large $t$ expansion (\ref{5sphereFinfty}). We are going to drop the exponentially small contributions proportional to powers of $e^{r/4g}$, since, as we will see, the phase transition occurs in a region where this is justified.\footnote{For a reference, at the triple point $e^{r/4g}$ is roughly $10^{-8}$.} These terms, including their contribution to the free energies and transition lines, can be computed and one needs to be a little careful in the region where $r/4g$ is not a very large number.\footnote{Up to second order the correction to the solution (\ref{solD}) read
\[\ldots+e^{\frac{r}{4 g}} \left(\frac{1}{4 g}-\frac{1}{4 g r}+\frac{1+8 g}{8 g r^2}-\frac{1+48 g}{24 g r^3}+\ldots\right)+e^{\frac{r}{2 g}} \left(\frac{-1+4 g}{16 g^2}+\frac{1-4 g}{8 g^2 r}+\frac{-1+8 g^2}{8 g^2 r^2}+\ldots\right)+\ldots\ .\]} That said, taking the appropriate limit in (\ref{4_cond2}) yields
\be
D=-\frac{r}{4 g}+\frac{1}{r}+\frac{4 g}{r^3}+\ldots
\ee
and then from (\ref{4_freeS2C}) the free energy
\be\label{freeS2C}
\F_{S2C}=-\frac{r^2}{16 g}+\frac{1}{2}\log\lr{-r}+\frac{3}{8}-\frac{1}{4}\log \lr{4g}-\frac{g}{r^2}-\frac{4 g^2}{r^4}+\ldots
\ee

\paragraph{Phase transition}
It is straightforward to Pade approximate all of these expressions in $1/r$ and check that they are in a good agreement with numerical solutions to corresponding equations obtained in \cite{jt18}.

Comparing the free energies (\ref{freeAS1C}) and (\ref{freeS2C}) yields the following condition for the phase transition
\begin{align}\label{5_trans_cond}
0=\frac{3}{8}+\log 2+\frac{1}{4}\log g-\frac{1}{8 r}+\frac{-1-72 g}{48 r^2}-\frac{1+276 g}{192 r^3}+\frac{-3-1400 g-46320 g^2}{1920 r^4}+\ldots
\end{align}
This condition is most naturally solved by expanding $g(r)$ in powers of $1/r$ and solving order by order. This yields the transition line
\be\label{3trafo_pert}
g=\frac{1}{16 e^{3/2}}+\frac{1}{32 e^{3/2} r}+\frac{9+5 e^{3/2}}{384 e^3 r^2}+\frac{141+16 e^{3/2}}{3072 e^3 r^3}+\frac{13545+18240 e^{3/2}+764 e^3}{368640 e^{9/2} r^4}+\ldots\ .
\ee
This is an alternating series in $1/r$, which allows for a nice Pade approximation
\be
g=\frac{\frac{1}{16 e^{3/2}}
+\frac{4095+4890 e^{3/2}+4 e^3}{640 e^{3/2} \left(162-243 e^{3/2}+2 e^3\right) r}+\frac{-92745+43920 e^{3/2}+4164 e^3+4 e^{9/2}}{7680 e^3 \left(162-243 e^{3/2}+2 e^3\right) r^2}+\ldots
}{1-\frac{3 \left(-285-3250 e^{3/2}+12 e^3\right)}{40 \left(162-243 e^{3/2}+2 e^3\right) r}+\frac{-121905+66330 e^{3/2}-30396 e^3+20 e^{9/2}}{480 e^{3/2} \left(162-243 e^{3/2}+2 e^3\right) r^2}+\ldots}\ .
\ee
This line has a limited range in $g$ and extends only to a finite value of $g=1/16 e^{3/2}$ as ${r\to-\infty}$ and the phase transition line has a vertical asymptote. This has already been observed in \cite{jt18}.

Note that it is straightforward to do the computation to a much higher order in $1/r$ than shown and such results will be used later to compute the triple point. We show this fourth order formula for brevity.

\subsection{Asymmetric one-cut to symmetric one-cut transition}\label{sec3.3}

\paragraph{Symmetric one-cut solution}
This is more tricky than the previous case. The reason is that the condition (\ref{4_cond1}) can not be reasonably treated in the ${r\to-\infty}$ limit. We however note that we can solve the equation (\ref{4_cond1}) at the phase transition point (\ref{5_line1}) so we are going to look for corrections to the eigenvalue density at this point.

In line with what we have done above, we drop the exponentially small terms, which are proportional to ${e^{-1/\sqrt g}}$ in this case, and write
\begin{align}
\delta & = \delta_0+ \delta_1 (r+5\sqrt{g}) + \delta_2 (r+5\sqrt{g})^2+\ .\ldots  
\end{align} 
From (\ref{4_cond1}) we rather straightforwardly obtain
\begin{align}
\delta & = \frac{2}{\sqrt{g}} - \frac{5\sqrt{g}+r}{3g}-\frac{(5\sqrt{g}+r)^2}{216g^{3/2}}-\frac{7 (5
\sqrt{g} + r)^3}{7776 g^2}+\ldots, 
\end{align}
with the expansion of the free energy:
\begin{align}
\F_{S1C} & = -\frac{9}{8}+\frac{1}{2}\log 2+ \frac{1}{2\sqrt{g}}(r+5\sqrt{g})-\frac{1}{12g}(r+5\sqrt{g})^2 +\ldots\ .\label{freeS1C}
\end{align}.

\paragraph{Phase transition}
The most technical part of the analysis comes with comparing the expressions (\ref{freeS1C}) with (\ref{freeAS1C}). They are both expansions around different values of $r$ and taking them at face value leads to little success. To make them understand each other we take the expansion (\ref{freeAS1C}) and use Pade approximants to complete the expansion and then do the expansion of this formula in powers of ${r+5\sqrt g}$. After this massage the two expressions are of the same character and we can look for the value $r$ that solves the condition ${\mathcal{F}_{sym}=\mathcal{F}_{asym}}$ order by order for a given value of $g$.\footnote{Note that it would be more natural to look for value of $g$ for a given value of $r$. However, since the expansion is in powers of ${r+5\sqrt g}$ this would still be an implicit equation.}

The final technical trick is to replace $g$ by a parameter $x$ given by
\begin{align*}
g= \frac{g_c}{(1+x)^2}\ ,
\end{align*}
where $g_c$ is the triple point.\footnote{Such $x$ is a better expansion parameter, as the transition line is then spanned by the values of $x \in (0, \infty)$. It also makes the expansion parameter $x$, rather than $\sqrt x$.} We will find the solution perturbatively in $x$:
\begin{align}
r= -5\sqrt{g_c}+ r_1x+r_2x^2+ r_3x^3+ r_4x^4+ \ldots\label{1cutrafo}
\end{align}
This way we obtain an uniluminating expression for $r$ which is essentially a power series in $1/\sqrt g$. We Pade approximate this formula in $x$, taking into account that we expect the transition line to pass trough the origin of the parameter space. We do not show the explicit formula for ecological reasons.

Note that the above procedure, in principle, solves the problem. However the condition determining $g_c$ is a transcendental equation that can not be solved analytically\footnote{Example of the condition for a very low order of expansion is
\be\label{0thORDER}\half \log 5+\frac{1}{4}\log g_c+\frac{5}{16}+\frac{3/4}{1+30 \sqrt{g_c}+120 g_c}=0\ .\ee}. We will therefore have to resort to the numerical solution of this one equation. Also note that the expressions determining $g_c$ and $r_i$'s involve all the terms from the small $t$ expansion of $F$, unlike the expansions in the Section \ref{sec3.2}. This means that going one order higher in the expansion (\ref{5sphereF0}) changes all the terms in the solution (\ref{1cutrafo}). The results however converge reasonably when increasing this order. An interesting observation is that the expansion of this approximation for small $g$ starts with a linear coefficient, rather than with $\sqrt g$ as does (\ref{5_line1}).

\subsection{Phase diagram and the triple point location}\label{sec3.4}
One way of obtaining the value of critical coupling is taking the numerical solution for $g_c$ from the previous section
\be\label{gcrit1}
g_c=0.004865\ .
\ee
To obtain an analytic value of the triple point without resorting to numerical solutions, we drop the exponentially small part in (\ref{5_line1}) and plug ${r=-5\sqrt{g}}$ in the phase transition condition (\ref{5_trans_cond}). This way, we obtain an equation for the value of the critical coupling ${g_c}$. This condition is then solved by introducing an auxiliary parameter $\ep$, keeping track of the order of terms in the original ${1/r}$ expansion. We thus have
\be 
0=\frac{3}{8}+\log 2+\frac{1}{4}\log g_c+\frac{\ep}{40 \sqrt{g_c}}-\ep^2 \left(\frac{1+72 g_c}{1200 g_c}\right)+\ldots\ .
\ee
and look for the solution order by order in $\ep$. Finally, we Pade approximate the resulting expression in powers of $\ep$ and set ${\ep=1}$. When working with expansion up to ${1/r^{20}}$ this yields for the critical coupling
\be\label{gcep}
g_c=0.0048655\ ,
\ee
which is reasonably close to the value obtained in \cite{jt18} by numerical solution of the transition conditions and agrees with (\ref{gcrit1}).

\begin{figure}
    \centering
    \includegraphics[width=0.8\textwidth]{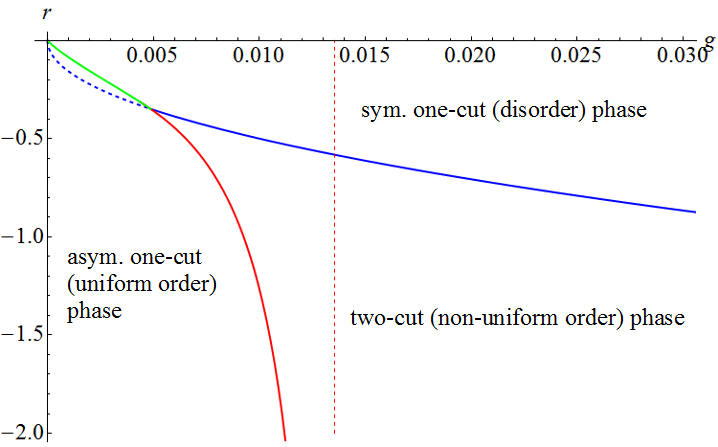}
    \caption{The phase diagram of the model with the effective action (\ref{5sphereF}). The green line separates the symmetric one-cut solution from the the asymmetric one-cut solution and is obtained using $1/r^{12}$ expansion, the blue line separates the symmetric one-cut and  the two-cut solutions and is exact, the red line separates the two-cut solution from the asymmetric one-cut solution and is computed using $1/r^{20}$ expansion. The vertical dashed line denotes the asymptote of the red line at ${g=1/16e^{3/2}}$.}
    \label{phasediag2}
\end{figure}

The whole phase diagram for this model is shown in the Figure \ref{phasediag2}. The asymptotic behaviour of the transition line between the two-cut and the asymmetric one-cut solution is at odds with the results of Monte Carlo simulations, which have identified a linear phase transition \cite{OConSaman}. This is a consequence of the approximation we have made neglecting the higher moment terms in the effective kinetic term action. For large and negative $r$ the two-cut configuration is far from the semicircle and the remainder term $\mathcal R$ in (\ref{4model}) gives a significant contribution.

\section{Modified kinetic term models}\label{sec4}
\subsection{General remarks}\label{sec4.1}

The purpose of this section is to show the path from the kinetic term $\K$ in (\ref{2fuzzyact}) to the effective action $F$ in (\ref{4model}), such as (\ref{5sphereF}). We will assume that the kinetic term has the same eigenfunctions as $C_2$, the polarization tensors $T_{lm}$.\footnote{They are fuzzy versions of spherical harmonics and ${C_2 T_{lm}=l(l+1)T_{lm}}$. Polarization tensors form a basis in the space of hermitian matrices \cite{bal}.} The form of the kinetic term is then given by a function $K(x)$ which determines the modified eigenvalues of $\K$ in the following way
\be \K T_{lm}=K\big(l(l+1)\big) T_{lm}\ .\ee
With a little abuse of notation we can write ${\K=K(C_2)}$.
Multiciplites of the eigenvalues are assumed to be $2l+1$ and we also rescale any parameters in $K$ such that the eigenvalues scale as $N^2$ in the large $N$ limit. Note, that one can handle the more general case of a fuzzy complex projective space $\mathbb C P^n$ in a very similar fashion \cite{steinacker05,jt15b}. The original eigenvalues would change to $l(l+n)$ and their multiplicity would be modified, but the rest of the analysis would straightforwardly follow.

The general procedure for obtaining the kinetic term effective action is as follows \cite{poly13}. One starts with the function (already with all the assumptions to accommodate the large $N$ limit)
\begin{align}\label{4_f}
    f(z)=\sum_{l=0}^{N-1}\frac{2l+1}{K\big(l(l+1)\big) +N^2 z }\to\int_0^1 dx \frac{1}{K(x)+ z}\ .
\end{align}
The derivative of the effective action is then given by the expression
\begin{align}\label{Fprime}
    F'(t)=\frac{1}{t}-f^{-1}(t)\ ,
\end{align}
which is the key equation. $f^{-1}(t)$ is the inverse function of the function (\ref{4_f}). The idea behind this procedure is to fix the effective coupling in the resulting multitrace model in a way that reproduces the known exact result for the eigenvalue distribution of the free theory model \cite{steinacker05}.
Equation (\ref{Fprime}) yields, together with condition ${F(0)=0}$,
\be\label{4Ffinal}
F(t)=\int_0^t d\tau\,\lr{\frac{1}{\tau}-f^{-1}(\tau)}=\int_{f^{-1}(t)}^\infty du\lr{u-\frac{1}{f(u)}} f'(u)\ .
\ee

As we have seen in the case of the fuzzy sphere, even if the function ${F(t)}$ is known explicitly, we can not hope for more than a perturbative solution of the equations determining the distributions and phase transitions. So in general all we can hope for is a perturbative solution to the determining equations and in principle all we need is the large and the small $t$ behaviour of $F$. Small $t$ expansion is going to be given by the large $z$ behaviour of $f(z)$ and vice versa.

Large $z$ expansion of ${f(z)}$ is straightforwardly given by
\be f_\infty(z)=\sum_{n=0}^\infty \frac{(-1)^n}{z^{n+1}}\int_0^1 dy\,K(y)^n=\frac{1}{z}+\frac{f_2^\infty}{z^2}+\ldots\ .\ee
We will use subscripts and superscripts $0$ and $\infty$ to denote objects relevant to the small and large argument expansion. We assume all the integrals to be finite in the above expression. Condition $f(z)=t$ can now be inverted order by order and (\ref{4Ffinal}) yields the small $t$ expansion of $F$ as
\be \label{Finfity} F_0(t)=\Big( -f_2^\infty\Big) t+\Big( f_3^\infty-\lr{f_2^\infty}^2\Big) t^2+\ldots\ .\ee

The small $z$ expansion of ${f(z)}$ is little more complicated to approach generally. For the kinetic term given by
\be 
K(x)=\alpha x+k(x)\ ,\ k(x)\textrm{ small and }\frac{k(x)}{x}\to0\textrm{ as }x\to 0\ ,
\ee
i.e. a small modification of the standard kinetic term, one can make some progress. We obtain
\begin{align}
   f_0(z)=&\sum_{n=0}^\infty (-1)^n\int_0^1 dx\, \frac{k(x)^n}{(\alpha x +z)^{n+1}}=\nonumber\\=&\frac{1}{\alpha}\log\lr{1+\frac{\alpha}{z}}+\frac{1}{\alpha}\sum_{n=1}^\infty(-1)^n\int_{1/(z+\alpha)}^{1/z}dt\, t^{n-1} k\lr{\frac{1/t-z}{\alpha}}^n\ . 
\end{align}
For polynomial $k(x)$, this is a series in $z$ including terms proportional to $\log z$ from integration of $1/t$ terms. The small $z$ expansion is going to have the following form
\be\label{4.F0} f_0(z)=-f^0_x(z)\log z+\sum_{n=0}^\infty f^0_n z^n\ee
and logarithmically diverge. This yields, after inverting order by order as before, the large $t$ behaviour of $F$ as
\be\label{3_Flarge} F_\infty(t)=\log t+C+F^\infty_1(t)e^{-t}+F^\infty_2(t)e^{-2t}+\ldots \ ,\ee
with no nice expressions for the coefficients $F^\infty$. The only large $t$ contribution to $F$ which is not exponentially suppressed is the part coming from the $1/t$ term in (\ref{Fprime}). If the second integral in (\ref{4Ffinal}) can be evaluated, the constant $C$ can be read of from the large $t$ expansion of $f^{-1}$. If this inverse is known exactly, the constant $C$ can be computed as follows
\be\label{4.C}
C=\lim_{\ep\to 0}\lr{-\log \ep-\int_\ep^\infty dt\,f^{-1}(t)}=\int_0^{f^{-1}(1)}du\,uf'(u)+\int_{f^{-1}(1)}^\infty du\,f'(u)\lr{u-\frac{1}{f(u)}}\ .
\ee
In the general case, computing this constant can be technically demanding and one may be forced to use some numerical or approximate methods when working with this formula.

Knowing the expansions (\ref{Finfity},\ref{3_Flarge}) is all we need to repeat the analysis of the Section \ref{sec3}. It would be very interesting to see what kinds of effective actions can be produced by different kinetic terms and what kinds of phase diagrams do different effective actions lead to. But as our main goal is in reasonable field theories, we will concentrate only on a modification of the kinetic term motivated by the scalar theory on the fuzzy sphere that is free of the UV/IR mixing.

\subsection{UV/IR mixing free theory}
Field theories on non-commutative spaces are non-local. As a consequence of this non-locality the short distance processes effect the large distance processes and the renormalization properties of the theories are spoiled. This exhibits itself as the UV/IR mixing phenomenon. The low momentum part of the non-planar Feynman diagrams of the theory diverges due to the UV divergence in the loop \cite{uvir1,uvir2}. In \cite{DolOConPres} it has been shown that the divergence of non-planar diagrams in the scalar field theory on the fuzzy sphere is limited to the tadpole diagrams. A particular modification of the kinetic term removes these diagrams form the theory and renders it free of the UV/IR mixing. In the terms of the matrix model, the modification is realized as follows
\be\label{UVIRfree}
S=\trl{\half M[L_i,[L_i,M]]+ 12 g M Q M+\half r M+g M^4}
\ee
where the operator $Q$ acts as
\be
Q T_{lm}=-\lr{\sum_{j=0}^{N-1} \frac{2j+1}{j(j+1)+r}\slr{N
(-1)^{l+j+N-1}
\left\lbrace\begin{array}{lll} l & s & s\\ j & s & s \end{array}\right\rbrace -1}}T_{lm}\ ,\ s=\half(N-1)\ ,
\ee
where $T_{lm}$ are again the eigenmatrices of $C_2$. It has been suggested in \cite{OConSaman} that as a first approximation we can disregard the $r$ dependence and consider the expansion of $Q$ in the powers of $l(l+1)$ 
\be\label{UVIRapprox}
\K=a C_2+ b C_2^2+\ldots
\ee
and take only the first two terms to study the modified theory (\ref{UVIRfree}). In what follows we will keep the values of $a$ and $b$ general.\footnote{We absorb the factor $12$ in (\ref{UVIRfree}) into the constants $a$ and $b$ in the rest of the text.}

This leads us to two most basic modifications of the standard kinetic term. First we will consider the ${b=0}$ case where the kinetic term is just rescaled by a coupling dependent factor. In the second case the kinetic term is modified by a term proportional to the square of the Casimir operator.

The UV/IR mixing has been argued to be the source of the striped phases in noncommutative field theories and thus it is natural to expect that the matrix models corresponding to the theories without this phenomenon will not support the two-cut solution. One way to see this is to observe that the kinetic term evaluated on the two-cut solution of the model is large and thus contributes significantly to the increase of the free energy. Since the model (\ref{UVIRapprox}) is just an approximation, we do not expect the phase to disappear completely, but expect some kind of a hint of the removal of the two-cut phase region. 

\subsection{Coupling enhanced fuzzy sphere model}\label{sec4.3}
The kinetic term is given by 
$\K M=(1+a g )\, C_2\,M$, with $a$ being a numerical factor, which we for the moment assume to be positive.\footnote{Note that $a$ needs to scale with the size of the matrix $M$ in order to keep the combination $a g$ finite.} This straightforwardly yields
\be\label{Fenh}
F[t]=N^2\log\lr{\frac{(1+a g) t}{1-e^{-(1+a g)t}}}\ .
\ee
One can check, by an explicit calculation using (\ref{4.C}) or by large $t$ expansion of the above expression that $C=\log\lr{1+a g}$.

There are two possible approaches to solving this model. We can repeat the analysis of the Section \ref{sec3} with this modified effective action and obtain modified conditions for the phase transitions. Or we can rescale the eigenvalues ${x_i\to \tilde x_i/\sqrt{1+a g}}$ to obtain a model with an effective action (\ref{5sphereF}) as in the Section \ref{sec3}, but with modified parameters\footnote{There is a constant shift to the free energy, but it is the same for all the solutions and does not modify the phase diagram.}
\be\label{tildeParameters}
\tilde r = \frac{r}{1+a g}\ \ \ , \ \ \ \tilde g=\frac{g}{(1+ag)^2}\ .
\ee
We can then use the final results obtained in the Section \ref{sec3} and translate them from the parameters ${\tilde r,\ \tilde g}$ to the parameters ${r,\ g}$. Both approaches lead to the same results, however the second is more instructive.

\subsubsection{Triple point}
The location of the triple point can be read directly from the equation
\be \tilde g_c=\frac{g_c}{(1+ag_c)^2}\ .\ee
This equation has two solutions
\be\label{sec4twosolutions} g_c=\frac{1-2 a \tilde g_c\mp \sqrt{1-4 a \tilde g_c}}{2 a^2 \tilde g_c}\ .\ee
The first solution is the new position of the triple point and moves away from the origin. The meaning of the second solution will be clear shortly, as well as the change of the behaviour at the value of the parameter $a=1/4 \tilde g_c$.

\subsubsection{Phase diagram}
Conditions (\ref{tildeParameters}) can be viewed as a deformation of the phase diagram in the Figure \ref{phasediag2}. However the relationship $g\leftrightarrow \tilde g$ is not monotonic. For small values of $g$ little changes qualitatively, but at the value ${g=1/a}$, or ${\tilde g=1/4a}$, direction of the deformation changes and beyond this point $g$ gets mapped onto small values of $\tilde g$ again.

\begin{figure}
    \centering
    \includegraphics[width=0.48\textwidth]{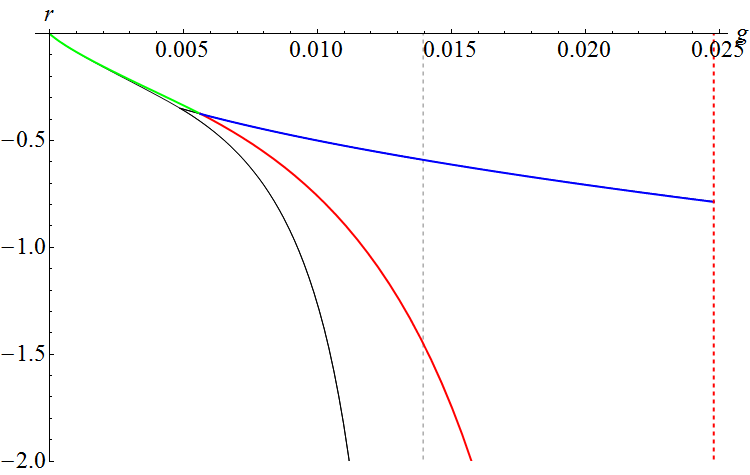}
    \includegraphics[width=0.48\textwidth]{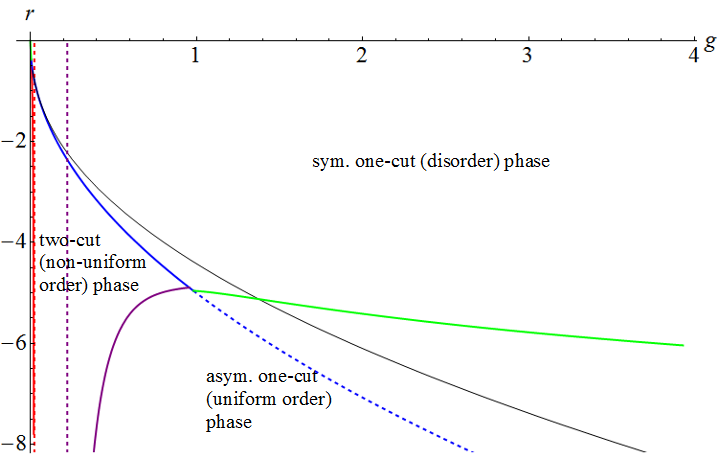}
    \caption{Two views of the phase diagram of the model with the effective action (\ref{Fenh}) for the value ${a=3 e^{3/2}\approx 13.45}$. The left image compares the original phase diagram (shown in thin black line) to the modified transition lines. As before, the green line separates the symmetric one-cut solution from the the asymmetric one-cut solution, the blue line separates the symmetric and one-cut and two-cut solutions and the red line separates the two-cut solution from the asymmetric one-cut solution. The right image shows the two new transition lines, purple between the two-cut and the new asymmetric one-cut solutions and green again between this new solution and the symmetric one-cut solution. In both images, the vertical dashed lines denote the asymptotes of the corresponding phase transition line.}
    \label{phasediagModified}
\end{figure}

This means that the part of the original diagram for ${\tilde g>1/4a}$ is not realized at all and for the values ${g>1/a}$ the same small $\tilde g$ part of the phase diagram gets repeated again, but backwards and deformed. This has huge consequences for the phase structure. There is a completely new region of parameter space, where the asymmetric phase is the preferred one. This region extends all the way to infinity in both $g$ and negative $r$ directions. The phase diagram of such model is given in the Figure \ref{phasediagModified}.

We can see that the triple point has shifted to the right, as did the asymptote of the asymmetric one-cut to symmetric two-cut phase transition. This is a hint of the mentioned receding of the two-cut phase, as expected with a model towards the removal of the UV/IR mixing. We also see the meaning of the second solution in (\ref{sec4twosolutions}). The scaling of the kinetic term with $g$ brought a completely new region, where the asymmetric solution dominates over the symmetric solutions (bottom right part of the diagram). It is however questionable, how much this is relevant for the original field theory model. As we have seen in the Section \ref{sec3} the second moment approximation does quite well for small values of the parameters, but fails for larger ones. We will therefore not dwell much further into consequences of this region for the field theory.\footnote{Since numerical simulations suggest that the asymmetric one-cut to symmetric two-cut phase transition line (the red line in all figures) is a straight line extending to infinity, one could argue that the second moment approximation underestimates the attraction brought in by the kinetic term and the situation is going to become only better in a more precise model. But we will not be overly optimistic here.}

The transformation of the phase diagram has a simple explanation in terms of the eigenvalue picture. The kinetic term introduces an attractive force among the eigenvalues. Scaling it with $g$ significantly enhances the attraction for large values of $g$ and at certain point this force wins against the eigenvalue repulsion from the Vandermode determinant even for deep wells of the potential. 

For a general value of $a$, this leads to three qualitatively different diagrams depending on the value of $a$:
\begin{itemize}
    \item ${0<a<4 e^{3/2}}$: The diagram consists of four regions as in the Figure \ref{phasediagModified} as described above.
    \item ${4 e^{3/2}<a<\frac{1}{4 \tilde g_c}}$: The two asymptotes merge together and the two-cut region does not extend all the way to the infinity. This happens because the phase transition condition does not have a solution in the leading $1/r$ order and the analysis of the previous section can not be repeated. One can Pade approximate directly the transition condition (\ref{5_trans_cond}) and try to solve it, but we will not go further into this here.
    \item $a>\frac{1}{4 \tilde g_c}$: There is no two-cut, or non-uniform order, phase. This presents a matrix model with a symmetric potential but with a complete spontaneous symmetry breaking of the $M\to - M$ symmetry.
    \item ${a<0}$: The trick (\ref{tildeParameters}) works only for ${g<-1/a}$. Beyond this point the overall $C_2$ coefficient is negative and the interaction brought by the kinetic terms is repulsive. This gives no hope for a stable asymmetric solution and even the one-cut solution needs much larger $r$ to exist due to the repulsion enhanced by $g$. However, as before, the approximation is not very relevant for large values of the parameters and we can say that for $a$ not too negative the triple point shifts to the left, as expected.
\end{itemize}

In conclusion we can say, that modification of the kinetic term by ${a g C_2}$ deforms the phase diagram in the Figure \ref{phasediagModified} and shifts it to the right for positive and to the left for negative $a$. The enhancement of the interaction by $g$ yields an interesting behaviour for larger values of $g$ and might hint at an mechanism of the removal of the two-cut phase in the UV/IR free theories, but remains questionable due to the approximation we have used.

To conclude this part, let us note that one could consider a simpler model without the coupling dependence in (\ref{Fenh}). This was originally proposed in \cite{OConSaman}. In this case, the transformation of the phase diagram as in the left image of the Figure \ref{phasediagModified} would be more substantial, since the small $g$ makes the modification less relevant. However there would be no new region of the asymmetric one-cut solution in the phase diagram and there would be no qualitative change from the original phase diagram.

\subsection{Higher derivative model}
A more substantial modification to the kinetic term is introduced by the second power of the Casimir operator
\be\label{4.4quadrakin}
\K M=(1+a g)\, C_2\,M+ b\,g\, C_2 C_2\,M\ .
\ee
We again assume that all the constants are scaled with $N$ such that all the terms in the action contribute in the large $N$ limit. Following the idea of the previous section, we rescale the eigenvalues by ${1/\sqrt{1+ag}}$, which results in a simpler model with ${\K=C_2+\tilde b C_2 C_2}$ and with parameters
\be\label{ABmodification} \tilde b=\frac{b\,g}{1+ag}\ ,\ \tilde r=\frac{r}{1+a g}\ ,\ \tilde g=\frac{g}{(1+a g)^2}\ .\ee
The plan is to map the solution of this model onto the model with the original parameters, similarly as we did in the previous section.

\subsubsection{Analysis of the simple model}
For such kinetic term $K(x)=x+\tilde b x^2$ and (\ref{4_f}) yields
\begin{align}
f(z)=&\frac{1}{\sqrt{1-4 \tilde b  z}}\log\left[\frac{1+2 z+\sqrt{1-4 \tilde b  z}}{1+2z-\sqrt{1-4 \tilde b z}}\right]=\\
\stackrel{z\to \infty}{=}&\frac{1}{z}
-\frac{3 + 2 \tilde b }{6 z^2}
+\frac{15 + 30 \tilde b  + 16\tilde b ^2}{180 z^3}
-\frac{-35 - 84 \tilde b - 70 \tilde b^2 - 20 \tilde b^3}{3780 z^4}
+\ldots\label{4.13}\\
\stackrel{z\to 0}{=}&\lr{-1-2\tilde b  z+\ldots}\log z-\log\lr{1+\tilde b }+\frac{1-2\tilde b ^2 -2\tilde b (1+\tilde b )\log \lr{1+\tilde b }}{1+\tilde b}z+\ldots\label{4.14}
\end{align}
and
\begin{align}
    f^{-1}(t)\,\stackrel{t\to 0}{=}\,&\frac{1}{e^{t}-1} \Big[1-\frac{\tilde b }{3}t+\frac{4 \tilde b^2 }{45}t^2+\frac{21 \tilde b+21 \tilde b^2-32 \tilde b^3 }{1890}t^3+\ldots\Big]\\
    f^{-1}(t)\,\stackrel{t\to \infty}{=}\,&\frac{1}{e^{t}-1} \Big[
   \frac{1}{1+\tilde b}+e^{- t} \frac{-2 \tilde b-3 \tilde b^2+2 \tilde b t+2 \tilde b^2 t}{(1+\tilde b)^3}
    +\nonumber\\+e^{-2 t}&\frac{-2 \tilde b-5 \tilde b^2+5 \tilde b^4+(4 \tilde b+2 \tilde b^2-14 \tilde b^3-12 \tilde b^4) t+(6 \tilde b^2+12 \tilde b^3+6 \tilde b^4) t^2}{(1+\tilde b)^5} +\ldots\Big]
\end{align}
where we have made use of the fact, that we can invert the ${\tilde b=0}$ case exactly. Using the second expression in (\ref{4Ffinal}) together with the above we obtain
\begin{align}
F(t)\stackrel{t\to 0}{=}&\frac{1}{6} (3+2 \tilde b ) t-\frac{16 \tilde b ^2 +30 \tilde b +15}{360} t^2+\frac{\tilde b (64 \tilde b ^2 +126 \tilde b +63) }{11340}t^3+\ldots \ ,\\
\stackrel{t\to \infty}{=}&\log t+\lr{1+\frac{1}{\tilde b}}\log\lr{1+\tilde b}-1+\frac{1}{1+\tilde b }e^{-t}+\frac{1+\tilde b+2 \tilde b t-\tilde b^2 (1-2 t)}{2(1+\tilde b)^3}e^{-2t} +\ldots\ ,
\end{align}

From this point on, the analysis of the model follows directly along the lines of Section \ref{sec3}.

The \emph{symmetric one-cut to symmetric two-cut} transition line can be calculated exactly and only the exponentially suppressed terms change for small values of $g$.

For the \emph{asymmetric one-cut to symmetric two-cut} phase transition we arrive at the following condition 
\begin{align}\label{4trafo_condition}
    0=&\frac{7}{8}+\log 2+\frac{1}{4}\log \tilde g-\half\lr{1+\frac{1}{\tilde b}}\log\lr{1+\tilde b}-\frac{3+2 \tilde b}{24\tilde r}+\no&+\frac{-10(1+72\tilde g)-15 \tilde b -6 \tilde b^2}{480\tilde r^2}+\ldots\ .
\end{align}
This is a deformation of (\ref{5_trans_cond}) and there is no qualitative change. We solve (\ref{4trafo_condition}) order by order in the powers of large and negative $r$ obtaining a modification of the solution (\ref{3trafo_pert})
\begin{align}
  \tilde g=&\frac{(1+\tilde b)^{2+\frac{2}{\tilde b}}}{16 e^{7/2}}+\frac{(1+\tilde b)^{2+\frac{2}{\tilde b}} (3+2 \tilde b)}{96 e^{7/2} \tilde r}+\nonumber\\&+\frac{(1+\tilde b)^{2+\frac{2}{\tilde b}} \left(135 (1+\tilde b)^{2+2/\tilde b}+(75 +105 \tilde b +38 \tilde b^2 )e^{7/2}\right)}{5760 e^7 \tilde r^2} +\ldots\ .
\end{align}
Pade approximation of this expression gives the final formula for the phase transition
\be\label{simpleQKtrafo}
\tilde g=\frac{\frac{(1+\tilde b)^{2+\frac{2}{\tilde b}}}{16 e^{7/2}}+\frac{(1+\tilde b)^{2+\frac{2}{\tilde b}} \left[15 \tilde b \left(-18 (1+\tilde b)^{2/\tilde b}+e^{7/2}\right)+15 \left(-9 (1+\tilde b)^{2/\tilde b}+e^{7/2}\right)+\tilde b^2 \left(-135 (1+\tilde b)^{2/\tilde b}+2 e^{7/2}\right)\right]}{960(3+2 \tilde b) e^7}\frac{1}{\tilde r}
+\ldots}{
1-\frac{15 \left(9 (1+\tilde b)^{2/\tilde b}+5 e^{7/2}\right)+15 \tilde b \left(18 (1+\tilde b)^{2/\tilde b}+7 e^{7/2}\right)+\tilde b^2 \left(135 (1+\tilde b)^{2/\tilde b}+38 e^{7/2}\right)}{60(3+2 \tilde b) e^{7/2}}\frac{1}{\tilde r}+
\ldots
}\ . \ee
Note that the whole procedure, starting with the computation of ${f(z)}$ is well defined only for ${\tilde b>-1}$. For smaller values the second moment approximation (\ref{4model}) is not well defined.

The determination of the \emph{asymmetric to symmetric one-cut} phase transition line follows Section \ref{sec3.3}, with the difference that the zeroth order condition (\ref{0thORDER}) is now a function of the parameter $\tilde b$ and thus can be solved numerically only for a particular value.

\begin{figure}
    \centering
    \includegraphics[width=0.48\textwidth]{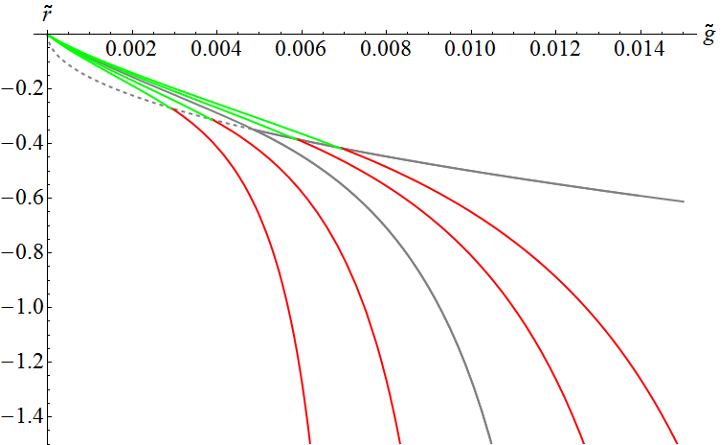}
    \includegraphics[width=0.48\textwidth]{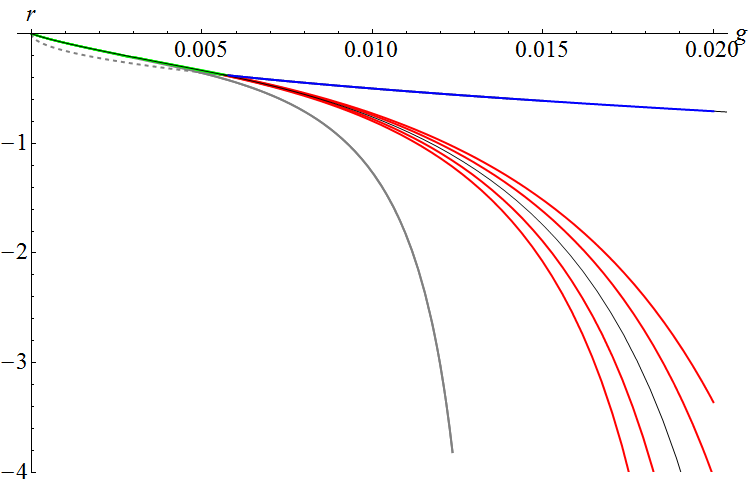}
    \caption{The right figure shows the phase transition lines of the quadrakinetic model (\ref{4.4quadrakin}) within the second moment approximation for ${a=0}$ and values of $\tilde b$ from left to right: ${-0.5,-0.25,0.25,0.5}$. The asymmetric one-cut to symmetric two-cut transition (\ref{simpleQKtrafo}) is shown in red, the asymmetric one-cut to symmetric one-cut in green. The symmetric phase transition is not changed for small values of $g$ and is the same as in the ${\tilde b=0}$ model diagram shown in gray.
    The left diagram is for the value ${a=3 e^{3/2}\approx 13.45}$, i.e. as in the Figure \ref{phasediagModified}, and the values of $b$ from left to right: ${-4,-2,2,4}$.}
    \label{phasediagQuadrakin}
\end{figure}

The resulting transition line together with transition line (\ref{simpleQKtrafo}) for several values of $\tilde b$ are shown in the Figure \ref{phasediagQuadrakin}. As expected, for negative values of the parameter the transition lines shift to the left, since the eigenvalue attraction from the $C_2$ term in the action is dampened by the $C_2^2$ term. The opposite is true for the positive values of $\tilde b$. Values of the triple points are given in the Table \ref{table1}. They have been obtained by the same procedure as the value (\ref{gcep}) in the Section \ref{sec3.4} for the simple fuzzy sphere model. The most important conclusion is that the properties of the phase diagram are not qualitatively different from the diagram for the simple fuzzy sphere model in the Figure \ref{phasediag2}.

\subsubsection{Analysis of the complete model}

\begin{table}
    \centering
    \begin{tabular}{ll}
    \begin{tabular}{r|l}
         $\tilde b$ & $\tilde g_c$ \\
         \hline
          $-0.5$ & $0.0029257$\\
          $-0.25$ & $0.0038845$\\
          $0$ & $0.0048655$\\
          $0.25$ & $0.0058787$\\
          $0.5$ & $0.0069285$
    \end{tabular}
    
    \hspace{10pt}
    \begin{tabular}{rr|l}
         $a$ & $b$ & $g_c$ \\
         \hline
          $3 e^{3/2}$ & $-4$ & $0.0055203$\\
          & $-2$ & $0.0055748$\\
          & $0$ & $0.0056300$\\
          & $2$ & $0.0056900$\\
          & $4$ & $0.0057233$
    \end{tabular}
    \end{tabular}
    \caption{Values of the triple points for the phase diagrams in the Figure \ref{phasediagQuadrakin}.}
    \label{table1}
\end{table}

We repeat the familiar drill. We use (\ref{ABmodification}) in (\ref{4trafo_condition}) and look for solution $g$ as a power series in ${1/r}$ to obtain the symmetric two-cut to asymmetric one-cut phase transition. We treat the symmetric one-cut to asymmetric one-cut phase transition of the simple model from the previous section to modification (\ref{ABmodification}) and repeat the procedure described in the Section \ref{sec3.3}. The zeroth order condition, which needs to be solved numerically, is again a function of the parameters $a$ and $b$ and thus we can obtain the complete transition line only for particular values of these. Large and small $g$ expansion of the symmetric one-cut to symmetric two-cut phase transition can be straightforwardly obtained from the explicit solution as in the Section \ref{sec3.1}.

Finally, we arrive at the phase diagram. As before, the approximation we are using is relevant for small values of $g$, so the properties of the phase diagram in this region are of the greatest interest. We will concentrate on the change of the small $g$ phase transition lines and we are not going to worry about the extra region which appears for larger values of $g$.

The resulting phase diagram for particular values of $a$ and $b$ is shown in the Figure \ref{phasediagQuadrakin} and the values of the triple point are given in the  Table \ref{table1}. These demonstrate the change in the phase diagram and in the location of the triple point. For positive values of $a$ and $b$ there are no issues in the phase diagram. If some of the values are negative, phase diagram can run into discrepancies for large values of $g$, but as long as the values $-1/a$ and $-1/(b+a)$ are larger than the important values of $g$ in the section \ref{sec3}, this is not relevant for the triple point location.

We thus conclude that the location of the triple point of the original diagram in the Figure \ref{phasediag2} can be controlled with the modification (\ref{4.4quadrakin}) and for positive $a$ and not too negative $b$ shifts to larger values of $g$, as expected from a modification towards the removal of the two-cut phase.

\section{Conclusions}\label{sec_conclusions}

There were two main goals to this work.

The first was to analyze the fuzzy sphere matrix model within the second moment approximation analytically, improving on the work \cite{jt18} where the equations determining the solution to the model have been solved numerically. We have been able to do so, identified the phase transition lines and computed the location of the triple point of the theory.

The second goal of this work was to investigate if the above method is capable of controlling the phase diagram also for theories with modified kinetic terms. We have shown that modification of the kinetic term does indeed lead to a modified phase diagram in this approximation and that the main features of the phase diagram, i.e. the existence of a stable asymmetric phase and of the triple point, are a general property of such models. We have also shown that modification of the kinetic term can control the location of the triple point, however the properties of the model did not change for large values of the parameters of the theory.

Thus the first line of a further research is in the analysis of the matrix models that go beyond the second moment approximation. The model analyzed in the section \ref{sec3} reproduces some of the features of the phase diagram of the fuzzy field theories, but fails for large values of the parameters. It would therefore be very interesting to go beyond the second moment approximation and to incorporate higher moments of the eigenvalue distribution into the probability distribution (\ref{4model}). The goal would be to have a model, which reproduces the features of the diagram for all values of the parameters. Namely the straight phase transition line between the uniform order and non-uniform order phases for large values and the triple point and the uniform order to disorder phases transition for small values. At the moment there is no such model. Recently, a solution to the correlation functions of a generic quartic matrix model has been presented \cite{gw19} and it would be very interesting to see if this sheds some more light into the models described here.

The second line of a further research is to repeat the presented analysis for different fuzzy spaces. Numerical results and analysis of perturbative models are available for fuzzy disc and three dimensional ${\mathbb R \times S_F^2}$, the second moment method is straightforwardly applicable to higher fuzzy $\mathbb C P^n$, where perturbative analysis is also available.

Finally, it would be very interesting to use such matrix models to calculate various different properties of fuzzy field theories. Numerical studies of correlation functions \cite{correlationFunctions1,correlationFunctions2} and entanglement entropy \cite{entanglement1,entanglement2,entanglement3} on the fuzzy sphere have been performed. All these works demonstrate very different behaviour of the fuzzy field theories from their commutative counterparts and matrix models could be a tool capable of investigating these differences analytically.

\acknowledgments

We are in debt to Denjoe O'Connor and Samuel Kov\'a\v cik for many fruitful discussions.

This work was supported by \emph{VEGA 1/0703/20} grant. The work of JT was supported by \emph{Alumni FMFI} foundation as a part of the \emph{N\'{a}vrat teoretikov} project, the work of M\v S has been supported by \emph{UK/432/2019} and \emph{UK/432/2018} grants. Both of the authors have greatly benefited from the support of the COST MP-1405 action \emph{Quantum structure of spacetime}. Authors would like to thank the \emph{STP DIAS} and organizers of the \emph{Corfu Summer Institute 2019} for hospitality during their visits.



\end{document}